\documentstyle[12pt,qqa4scsk,psfig]{article}
\input tcilatex

\QQQ{Language}{
American English
}

\begin{document}

\author{P. N. Bhat  \\ 
{\it Tata Institute of Fundamental Research, }\\{\it Homi Bhabha Road,
Mumbai 400 005, India.}}
\title{THE ROLE OF FLUCTUATIONS IN ATMOSPHERIC \v CERENKOV TECHNIQUE}
\maketitle

\begin{abstract}
In the absence of a standard source of gamma rays or hadrons of known energy
one has to study the details of production of \v Cerenkov light at the
observation level only through detailed simulation studies. Recently such
studies have become all the more important in view of the various techniques
resulting from such studies, to distinguish between gamma ray initiated
events from those generated by much more abundant hadronic component of
cosmic rays. We have carried out a detailed simulation studies using the
CORSIKA package in order to study the \v Cerenkov photon density
fluctuations at various core distances both for photon and proton primaries
incident vertically at the top of the atmosphere. It is found that the
density fluctuations are significantly non-statistical. Such fluctuations
are much more pronounced in the case of proton primaries as compared to
photon primaries at all energies. Several statistical parameters have been
computed some which might lead to a technique in distinguishing photon
primaries.
\end{abstract}

\section{INTRODUCTION}

Ground based atmospheric \v Cerenkov technique is the only way by which TeV $%
\gamma $-rays could be detected from point sources such as $\gamma -$ray
pulsars, short period X-ray binaries or BL-Lac objects. Recent detection of
TeV emission from a few of these objects (Vacanti et al., 1991; Punch, et
al., 1992) has created much interest in the field of TeV $\gamma -$ray
astronomy. The $\gamma -$ray signals found typically are $\sim 1\%$ of the
abundant background events of cosmic ray nuclei, particularly protons. In
order to detect faint Very High Energy (VHE) $\gamma -$ray sources, one has
to enrich the data with the signal by rejecting a bulk of the hadronic
background. In order to do so it is imperative to study the detailed
characteristics of \v Cerenkov light production by photon initiated and
proton initiated cascades in the atmosphere.

We have, therefore planned to carry out detailed simulation studies using
the CORSIKA (Knapp and Heck, 1996)\ package for various photon and proton
energies incident on the top of the atmosphere. The input parameters used in
the package correspond to that of Pachmarhi Array of \v Cerenkov Telescopes
(PACT, Bhat et al., 1995). Eventually, these studies are expected to lead to
the derivation of an event parameter which would enable us to distinguish
between the two types of primaries. In this paper we present preliminary
results from such studies involving the \v Cerenkov photon density
fluctuations at different primary energies as a function of core distance.

\section{THE ARRAY}

PACT, described elsewhere (Bhat et al., 1995; Bhat, 1996) consists of 25 \v
Cerenkov telescopes deployed in the form of a $5\times 5$ array over an area
of 80 m$\times $100 m located at Pachmarhi (latitude: $22^{\circ }28^{\prime
}N$ longitude: $78^{\circ }26^{\prime }E$). Each \v Cerenkov telescope
consists of 7 parabolic reflectors making up a total area of 4.45 $m^2$.
Individual phototubes are placed at the focus allowing us to carry out
multiple sampling of photons at a telescope. The expected energy threshold
of an event triggering the computerised data acquisition system is expected
to be around 350 GeV. We plan to distinguish between the photon initiated
showers from those of the proton initiated showers by the measurement of
photon densities received at each telescope. In addition, the fine timing
measurements will enable us to measure the shower direction correct to $\sim
0.2^{\circ }$ because of which we can reject the off-axis showers which are
presumably initiated by hadrons.

\section{THE SIMULATIONS}

About 100 showers have been generated for vertically incident $\gamma -$ray
photons and protons at the top of the atmosphere. The number of \v Cerenkov
photons in the wavelength range 300-450 nm incident at each telescope is
written out on the data file. Several such simulation studies have been
carried out in the past (Hillas and Patterson, 1990; Rao and Sinha, 1988;
Hillas, 1982). Initially it was decided to study the role of density
fluctuations at various core distances. The question one is trying to answer
is how do the fluctuations depend on the nature of the primary and how does
it depend on the core distance. For this purpose the frequency distributions
were derived at each detector and fitted to a known probability distribution
function. The statistical parameters like various moments were used as a
measure of the nature of these fluctuations.

\section{RESULTS}

\begin{figure}
\centerline{\psfig{figure=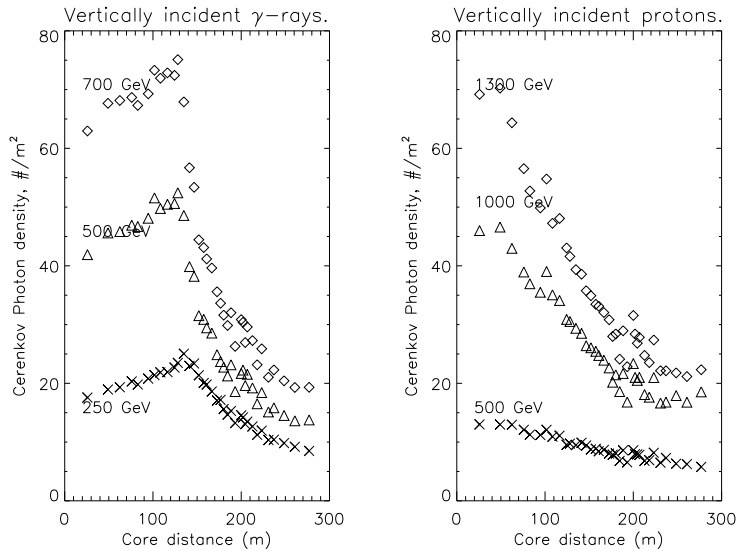,height=7.5cm,angle=0}}
\caption{\it Average lateral distribution of \v Cerenkov photon densities generated by 
$\gamma$-rays (at 3 different energies 250, 500 \& 700 GeV) and protons (at 3 different energies 500 GeV, 1000 GeV \& 1300 GeV with comparable \v Cerenkov yields) incident vertically at the top of the
atmosphere.}
\label{Figure 1:}
\end{figure}
Figure 1 shows the average (100 showers) lateral distributions derived for
proton and photon primaries with comparable \v Cerenkov yields. As expected
the photon showers are characterized by the increased density at a core
distance of $\sim 120m$, which is commonly known as the `{\it hump}' region,
at all the three primary photon energies viz. 250, 500 and 700 GeV. Figure 2
shows the variation of the various statistical parameters as a function core
distance at a fixed energy of 250 GeV for $\gamma -$ray primaries and 500
GeV for protons. On the top left corner the variation of the ratio of the
variance to the mean is shown which shows a clear separation for $\gamma -$%
ray and proton primaries. It may be noted that this parameter is rather
large even for $\gamma -$ray primaries while it is expected to be unity for
a Poisson distribution, thus showing that the density fluctuations are
significantly non-statistical consistent with previous studies (Sinha,
1995). The plot on the top right shows normalized $\chi ^2$ values for a
Gaussian fit to the density distribution. While distribution for a $\gamma -$%
ray primary fits well with a Gaussian (when the RMS is used as the standard
deviation) it does so poorly for proton primaries. The other two plots show
similar variations for third and fourth moments.

\begin{figure}
\centerline{\psfig{figure=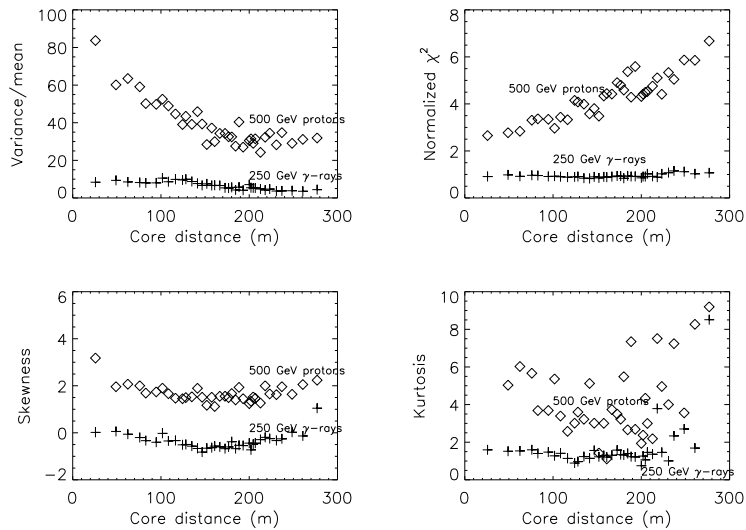,height=7.5cm,angle=0}}
\caption{\it Variation of 4 statistical parameters derived from \v Cerenkov photon
density distributions as a function of the core distance, for 250 GeV 
$\gamma$-ray and 500 GeV proton primaries.  
}
\label{Figure 2:}
\end{figure}
\section{CONCLUSIONS}

From our present simulation studies it is found that the \v Cerenkov photon
density fluctuations, measured as a ratios of the variance to the mean, are
highly non-statistical at all core distances. These ratios are much larger
in the case of proton primaries than for $\gamma -$ray primaries at all core
distances even though the difference seems to reduce at higher primary
energies. For a given primary, this ratio seems to be proportional to the
mean density itself. The density distribution at a given core distance for a 
$\gamma -$ray primary seems to fit with a Gaussian if one uses the RMS as
the standard deviation while that for proton primary does not fit well even
though the fit improves at higher primary energies. The third and fourth
moments show that the density distributions for proton primaries are more
asymmetric and more uneven compared to that of $\gamma -$ray primaries while
these differences reduce with increasing primary energies.

\section{REFERENCES}

\begin{description}
\item  Bhat, P. N., Acharya, B. S., Krishnaswamy, M. R. et al. in {\it %
Towards a Major Atmospheric \ Cerenkov Detector-IV, }ed. M. Cresti, pp.
194-201, (1995)

\item  Bhat, P. N. in {\it Perspectives in High Energy Astronomy \&
Astrophysics, }ed. P. C. Agrawal, (1996)

\item  Hillas, A. M., {\it J. Phys. G: Nucl. Phys.}, 8, 1475, (1982)

\item  Hillas, A. M. and Patterson, J. R., {\it J. Phys. G: Nucl. Phys.}, 16,
1271, (1990)

\item  Knapp, J. and Heck, D., {\it EAS Simulation with CORSIKA, V4.502: A
User's Manual}, (1996)

\item  Punch, M., et al., {\it Nature}, 358, 477 (1992)

\item  Rao, M. V. S. and Sinha, S., {\it Phys. G: Nucl. Phys.},14, 811, (1988)

\item  Sinha, S., {\it J. Phys. G: Nucl. Phys.}, 21, 473, (1995)

\item  Vacanti , G., et al., {\it Ap J}, 377, 467, (1991)
\end{description}

\end{document}